\documentclass[twocolumn,twoside,slac_two]{revtex4}
\usepackage{graphicx}
\usepackage{subfigure}
\usepackage{fancyhdr}
\begin{document}

\title{A needlet-based approach to the full-sky data analysis}
\author{R. Iuppa}
\affiliation{Department of Physics, University of Tor Vergata and INFN, Sezione di Roma Tor Vergata, via della Ricerca Scientifica 1,  00133 Roma , Italy}
\begin{abstract}
In cosmic-ray physics, large field of view experiments are triggered by a number of signals laying on different angular scales: point-like and extended gamma-ray sources, diffuse emissions, as well as large and intermediate scale cosmic-ray anisotropies. The separation of all these contributions is crucial, mostly when they overlap with each other. 
Needlets are a form of spherical wavelets that have recently drawn a lot of attention in the cosmological literature, especially in connection with the analysis of CMB data. Needlets enjoy a number of important statistical and numerical properties which suggest that they can be very effective in handling cosmic-ray and gamma-ray data analysis.
An application of needlets to astroparticle physics is shown here. In particular, light will be thrown on how useful they might be for estimating background and foreground contributions. Since such an estimation is expected to be optimal or nearly-optimal in a well-defined mathematical sense, needlets turn out to be a powerful method for unbiased point-source detections.
In this paper needlets were applied to two distinct simulated datasets, for satellite and EAS array experiments, both large field of view telescopes. Results will be compared to those achievable with standard analysis tecniques in any of these cases.
\end{abstract}

\maketitle

\thispagestyle{fancy}


\section*{Introduction}
Over the last decades very high energy cosmic-ray experiments made great headway in sensitivity and duty-cycle, providing huge amounts of very high quality data. At the present time space telescopes are operated with angular resolution as good as few tens of arcminutes. Large field of view ground-based telescopes easily achieve $1\deg$ capability in reconstructing the primary arrival direction. Every day experimenters cope with problems like resolving sources close to each other or separating them from diffuse or extended back/foreground or enhancing the significance of the detection, as well as defining the shape of extended sources or giving accurate estimation of directional anisotropies. For instance, ARGO-YBJ observed very intense localized excesses of cosmic rays, $10^\circ-30^\circ$ wide \cite{icrc0507}. The contribution of these regions must be carefully considered  when looking at gamma-ray sources nearby and viceversa. The Cygnus region may present similar problems, because many known gamma-ray sources are there within few degrees. The case of crowded regions is quite common in satellite experiments as sensitive as Fermi-LAT, mostly when parts of the Galactic plane are considered (think of the Vela region or the Galctic center).

In the framework of the standard analysis techniques, if the density in a certain direction on the sphere is considered \emph{as it is}, no information can be obtained about the angular scales which the signal comes from in that point. To get \emph{exactly} such information, the harmonic expansion can be used, but in this case no localization in the real space can be obtained anymore. A very good tradeoff is represented by the wavelet expansion, where the exactness of the harmonic expansion is given up in favour of the possibility of having good localization properties in real and harmonic domain at the same time. This is the reason why a growing interest has been devoted in the last five years to the application in a cosmological environment of a new form of spherical wavelets, called {\em needlets}. Needlets were introduced in the mathematical literature by \cite{npw1}, see also \cite{gm1} for extensions and generalizations;  several applications to Cosmic Microwave Background data data have also been implemented: see for instance \cite{mpbb08,scodeller}. More recently, a few papers have focussed on the use of needlets to develop estimators within the thresholding paradigm, in the framework of directional data, which provide the mathematical formalism for cosmic rays experiments \cite{bkmpAoSb}.

In this paper, we shall first review briefly the main features of needlets and the explain how their properties make them a very promising tool for cosmic rays data analysis. To point out how effective they may be either for satellite or ground experiments, needlets are applied here to two distinct simulated datasets, which reproduce the data acquisition of Fermi-LAT and ARGO-YBJ. The second section is then aimed at describing the simulation. Afterwards, we present applications of needlet procedures on the simulated dataset. A final section summarizes results and discusses perspectives for further research.
\section{Needlets construction and main properties}
\label{sec:needlets}
Needlets enjoy quite a few important properties that make them very suitable for spherical data analysis. Indeed, needlets do not rely on any tangent plane approximation and they are perfectly adapted to standard packages; their main property is the capability of unfolding directional data in the harmonic space, so that every order $j$ of needlets contains power only from within a certain range in the $l$ space. In the real space for any fixed angular distance the tail of the needlets decay faster than any polynomial, i.e. quasi-exponentially as the frequency increases. It is worth noticing that for needlets the exact reconstruction property holds, i.e. back-transform exists giving the input signal within numerical accuracy. 

We do not give details about needlets construction here (on this purpose, see \cite{npw1,bookmarinucci}). Nonetheless, we recall that the needlet idea has been extended by Geller and Mayeli \cite{gm1} with the construction of so called Mexican needlets, see  also \cite{scodeller} for numerical analysis and implementation in a  cosmological framework. Mexican needlets have localization properties in real space even better than standard needlets, that is why they were applied to get the results presented here.

Numerical codes have been developed by the author to compute the needlet estimators $\widehat{\beta}_{jk}$ for any sky map containing density information. Some other codes are publicly available \cite{needatool} and results reported here have been suitably cross-checked. The index $j$ refers to the needlet order, while $k$ runs along the pixels into which the sphere has been discretized (the Healpix package is used here \cite{Healpix}, with the ``ring'' numbering scheme).

When the needlet transform is numerically computed, three parameters are important, i.e. $j_0$ (the first order computed) and $n_j$ (how many needlet orders are computed) and $B>1$ which is a number used to pass from the $j$-th order to the $(j+1)$-th. A golden rule to be used is that the order $j$ contains power mostly from multipoles in the range $[B^{j-1},B^{j+1}]$\footnote{That is, if $B=1.6$, then $j=6$ corresponds to $l\sim1.6^{6}=17$, or $\psi\sim10^{\circ}$.}. If mexican needlets are used, one more parameter is important, $p$, which describes how fast needlets go to zero in the harmonic space.

We shall now consider the analysis of data from cosmic rays observatories, following classical approaches to wavelet-based density estimation. An idea that we shall pursue is to implement \emph{thresholding} estimates, as discussed for instance by \cite{bkmpAoSb}. More precisely, we can consider the nonlinear estimate
$\widehat{f}_{n}^{\ast }(x)$ obtained by suitably weighting the needlet coefficients in the back-transform. The weight function $w_{jk}(\widehat{\beta }_{jk})$ is some nonlinear function that ``shrinks'' beta towards zero. Such estimates can be shown to be robust and nearly optimal over a wide class of density functions and different loss functions, i.e. figures of merit by which to measure when the estimate is "close" to the density to be estimated.

Hereafter, we will name {\em source-map} the sky-map as it comes from the experiment; {\em beta-maps} those containing the coefficient estimators $\widehat\beta_{jk}$; {\em reconstructed maps} or {\em back-transformed maps}, those obtained after the application of the inverse needlet transform.
\section{Needlet orders and angular scales}
As for any mathematical transform, the effectiveness of needlets lies in the chances that they offer to the analyst when handling data in the transform space. In this sense, it is crucial to understand how certain given signals are represented in the transform space. Even more, in which part of the transform space the highest power of the signal is saved. On this purpose, the golden rule outlined in the previous section is an important basic tool. More in detail, many figures of merit can be defined to characterize the performance of the needlet transform, mostly looking at the reconstructed signal. Just to give an example, we introduce here two quantities which can be evaluated for simulated signal from point-like sources:
\begin{itemize}
\item $\rho_r=N_{r}^{R}/N_{r}^{S}$, where $r$ is a certain angular distance from the source and $R$ and $S$ stand for ``reconstructed'' and ``sampled'' respectively. $\rho_r$ is the ratio of the number of events reconstructed within $r$ from the source to the number of events sampled in the same region;
\item $\langle\delta_{r}\rangle=\sqrt{\sum_{i=1}^{n_r}\,\left[({\mathcal N}_{i}^{R}-{\mathcal N}_{i}^{S})/{\mathcal N}_{i}^{S}\right]^2/n_r}$, where the bin index $i$ takes values $[1,n_r]$ corresponding to angular distances from the source $(0,r)$. ${\mathcal N}_{i}^{R}$ and ${\mathcal N}_{i}^{S}$ indicate the number of events in the $i^{th}$ bin for the ``reconstructed'' and ``sampled'' map respectively. $\langle\delta_r\rangle$ is the square root of the $\chi^2$ of the reconstructed radial distribution to the sampled one, after normalizing it to the bin number. Smaller $\langle\delta_r\rangle$, higher the confidence that not only the integral, but also the shape of the reconstructed signal reproduces well the sampled one. 
\end{itemize}
To properly characterize the method, we computed $\rho_r$ and $\langle\delta_r\rangle$ for many combinations of parameters $(B,j_0,n_j)$, for both standard and Mexican needlet, as well as for many source extensions and intensities. Many detector point spread functions (PSFs) were accounted for, too.

\begin{figure}[!h]
  \centering
  \includegraphics[width=0.45\textwidth]{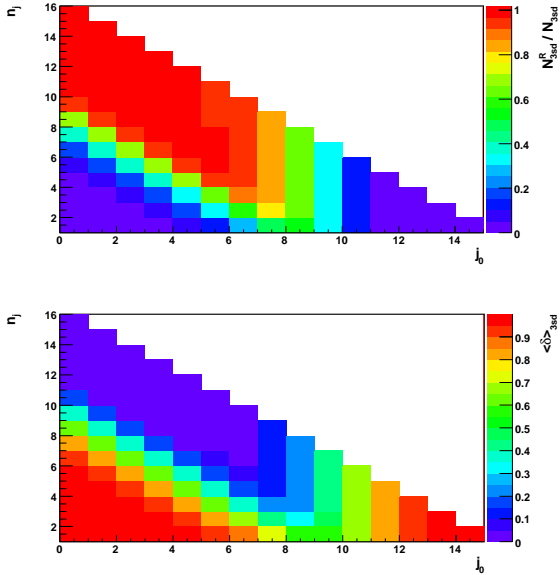}
  \caption{Representation of the reconstruction properties of the Mexican needlet transform. {\em Upper plot:} $\rho_{3\sigma}$. {\em Lower plot:} $\langle\delta_{3\sigma}\rangle$. See text for definition and details.}
  \label{fig:schemes}
\end{figure}

By way of example, figure \ref{fig:schemes} reports schematic views of $\rho_r$ and $\langle\delta_r\rangle$ as functions of $j_0$ and $n_j$. To realize it, we simulated a source emitting $10^6$ events, assuming a gaussian PSF with $\sigma=1^\circ$ and applied the Mexican needlet transform $(B=1.6,\,p=1,\,j_0=1,\,n_j=15)$. To oversimplify the example, no fore/background contributions are superimposed. The radius chosen to evaluate $\rho$ and $\langle\delta\rangle$ was $r=3\sigma$ and no background contribution is taken in consideration. For both plots, the first order used to reconstruct the signal is represented on the horizontal axis, while the vertical axis number of orders used. The color scale represents the values of $\rho_{3\sigma}$ and $\langle\delta_{3\sigma}\rangle$. For instance, bins $(5,6)$ contain the values of $\rho_{3\sigma}$ and $\langle\delta_{3\sigma}\rangle$ obtained from the map reconstructed with needlet orders $5\rightarrow10$.
What is easy noticing is that the two plot are somehow conjugate. Where the ratio $\rho_{3\sigma}$ is close to 1, the error $\langle\delta_{3\sigma}\rangle$ is close to 0 and viceversa. It happens in the top-left part of the schemes, representing cases where almost all the beta-maps computed in the transform are used in the reconstruction. Nonetheless a very good reconstruction can be obtained also with less expensive choices: e.g. if orders $5-11$ are used, more than $95\%$ of events will be part of the reconstruction and the signal shape will be good too, because the error is less than $5\%$.  
\section{$\beta$-maps and significance estimation}
To fully exploit the potential of needlets, analysis can be carried out in the beta space, i.e. handling beta-maps. Selections and operations can be made there, after which data may be backtansformed if needed. As often in the real space the analysis is performed on the basis of what the experimenter expects in case the signal is not there (null detection hypothesis), all the same knowing how beta-maps would look like in absence of signal is essential.

If there exists a reliable way of simulating how the sky would appear without the signal under study, an estimation of the average and the r.m.s. of the $\widehat{\beta}_{jk}$ is easily achievable, what can be used as reference in managing the data in the beta space. This method gives an \emph{ad-hoc} estimation for every pixel, although it inherits all the systematic uncertainties from the Montecarlo simulation. Results reported in the section \ref{sec:argo_fermi} are obtained in this way.

Simulation is the main way, but not the only one. If what the experimenter considers noise \footnote{Whichever angular scale it belongs to.} is uniformly distributed in the sky region under consideration, then the $\widehat{\beta}_{jk}$ average and r.m.s. of may be estimated in a small peripheral corner. This technique has the advantage to estimate the background directly from data, but values obtained are the same for all pixels considered. Results reported in the section \ref{sec:argo_fermi} are obtained in this way.

Once the variances $\sigma^2_{jk}=\langle(\widehat{\beta}_{jk}-\langle\widehat{\beta}_{jk}\rangle)^2\rangle$ are obtained, they can be used to evaluate the significance of the observed signal in the beta-space, i.e. to implement the detection procedure \emph{directly} in the real space.
\section{Thresholding}
For the most part, analysis coincides with selecting only those $\widehat{\beta}_{jk}$ which are above a certain threshold (\emph{thresholding}). Such threshold can be defined on the basis of local criteria (e.g. pixels are kept only that are within $\psi$ from the considered region, \emph{block thresholding}), as well as statistical hypotheses. For instance, the variance-maps $\sigma^2_{jk}$, however estimated, can be used as term of comparison, to get rid of the non-significant coefficients. We define here a \emph{soft thresholding} technique whose results will be presented in the next sections. Let us introduce the weight functions $$w_{jk}(\widehat{\beta}_{jk})=\frac{1}{2}\left(1+\tanh\frac{\widehat{\beta}_{jk}/\sigma_{jk}-T}{L}\right)$$ where $T$ is a certain threshold, passing which the weight function goes from $0$ to $1$. The $L$ parameter is related to the width of the transition region.
\section{ARGO-YBJ and Fermi-LAT simulation}
\label{sec:argo_fermi}
The first application of needlets to real high energy astrophysics occurred for the ARGO-YBJ experiment \cite{icrc0518}. ARGO-YBJ is an air shower array able to detect the cosmic radiation at an energy threshold of a few hundred GeV. Here we reproduce the analysis reported there, by simulating data from three years of data acquisition of the ARGO-YBJ detector. The trigger rate, the duty-cycle, the field of view and the angular resolution have been suitably reproduced. Charged cosmic rays have been simulated
isotropically, according to what reported in \cite{ArgoMoon}; intermediate scale anisotropies have been added according to \cite{icrc0507}. As far as TeV gamma-ray sources are concerned, the Crab nebula, Mrk421, Mrk501, MGRO1908 and the Cygnus region were included in respect of the average spectra reported in the literature. No large scale anisotropies is there.

To give an idea of what the method might obtain if applied to satellite-borne experiments, needlets have been applied also on a simulation of 1 year of data taking of Fermi-LAT. All 1451 sources present in the 1FGL catalogue have been simulated. The galactic diffuse emission has been simulated using a Fermi collaboration template\footnote{gll\_iem\_v02.fit.} and the extragalactic diffuse emission has been sampled with spectral index -2.4. Neither electrons or albedo photons have been accounted for.

There is some oversimplification in neglecting the detector response function, as well as in considering the exposure to be uniform all over the sky. No deep study of denoising properties have been performed yet, then this work has to be considered as preliminary with respect to other similar wavelet applications to cosmic ray physics (see for instance \cite{starckps}).
\section{Results for the ARGO-YBJ simulation}
\begin{figure}[!h]
  \centering
  \includegraphics[angle=90,width=0.4\textwidth]{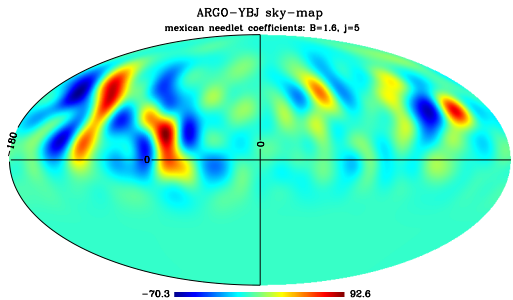}
  \caption{Beta-map from the simulated ARGO-YBJ sky. Mexican needlet transformation with $B=1.6$, $p=1$, $j=5$. The color scale indicates the intensity of the beta coefficients. Mollweide projection, equatorial coordinates.}
  \label{fig:beta}
\end{figure}

The figure \ref{fig:beta} represents the beta-map obtained for the ARGO-YBJ simulated map, order 5. As we expect from the corresponding angular scale ($(1.6)^5\sim17^\circ$), intermediate scale anisotropies \cite{icrc0507} are well visible. 

\begin{figure}[!h]
  \centering
  \includegraphics[width=0.4\textwidth]{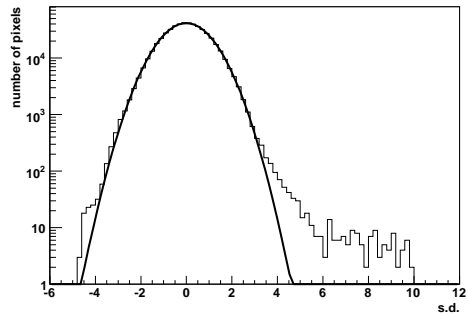}
  \caption{Significance distribution of the pixels of the $9^{th}$ order beta-map from the simulated ARGO-YBJ sky.}
  \label{fig:sigma}
\end{figure}

To evaluate the significance of the signal with respect to the estimated background, we used an independent background estimation ${\mathcal B}_k$ as mean map and sampled 1000 random background copies ${\mathcal B}^i_k{\rm,\ i=1,\dots,1000}$; every map has been needlet-transformed to compute the variance $\sigma_{jk}^2$ These maps were directly used to calculate the significance of the content of the $k^{th}$ bin in the $j^{th}$ beta-map: $\beta_{jk}/\sigma_{jk}$. For instance, the figure \ref{fig:sigma} shows the distribution of $\beta_{9k}/\sigma_{9k}$.  
The curve represents the gaussian fit to the $\widehat\beta_{9k}/\sigma_{9k}$ distribution, where $\hat\beta_{9k}$ is the transform of a background random copy (background subtracted). The fit results are $\mu=-0.0005\pm0.0011$ and $\sigma=1.0014\pm0.0021$, as it should be ($\chi^2/{\rm d.o.f.}=38.9/37$). The distribution of the signal map significantly deviates from the gaussian trend. The contribution of the sources is well visible from 3.5 up to 10 s.d. There is also a small excess below $-3$ s.d., whose origin is due to the shape of the needlet function.

As anticipated in the previous sections, this method allows to estimate the significance of every excess {\em directly} in the $\beta$ space, with no need of signal-smoothing or averaging. Moreover, it must be recalled that a given signal may be there in two or more $j$ orders and that the significance estimated here refers to the {\em single-order}.

\begin{figure}[!h]
  \centering
  \includegraphics[width=0.3\textwidth]{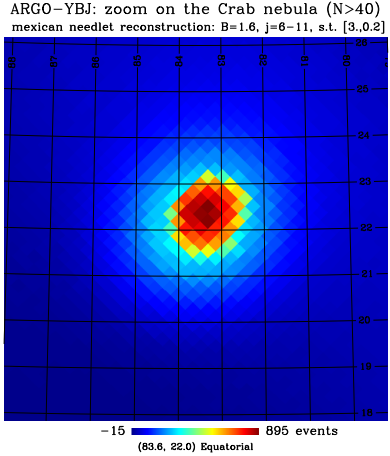}
  \caption{Signal of the Crab region reconstructed with the Mexican needlet technique. Parameters of the transform: $B=1.6$, $p=1$, $j=5-11$, soft thresholding $[3.0,0.2]$}
  \label{fig:crab}
\end{figure}

Finally it comes an example of application of the ``soft thresholding'' method. We set $T=3$ and $L=0.2$ for orders $j=5-11$\footnote{We choose these orders because of the PSF of ARGO-YBJ for point-like gamma-ray sources at $N>40$.} of the Mexican needlet transform ($B=1.6$ and $p=1$), then got the reconstructed map. These choices are optimized for point-like sources. A zoom of the result around the Crab nebula is given in figure \ref{fig:crab}. 
The source is well visible and no fluctuations typical of images from EAS arrays are there. The number of events reconstructed within $5^\circ$ from the nominal source position is $(114\pm3)10^3$, to be compared with that obtained from standard analysis techniques (e.g. \cite{mrk421paper}) $(116\pm2)10^3$ and that simulated $115234$.
\section{Results for the Fermi-LAT simulation}
\begin{figure}[!h]
  \centering
  \subfigure[]{\includegraphics[angle=90,width=0.4\textwidth]{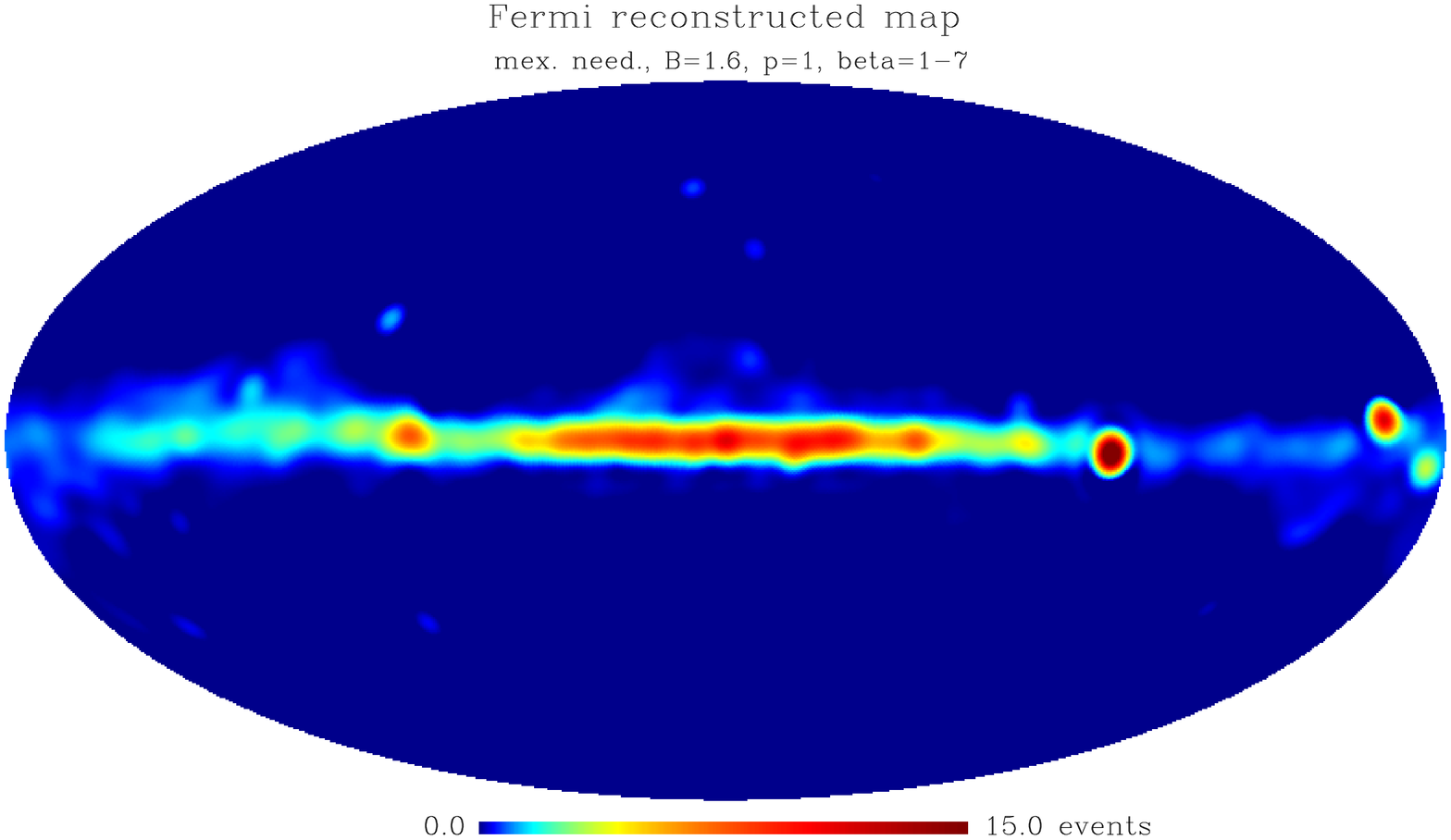} \label{fig:fermi_recon_1-7}}
  \subfigure[]{\includegraphics[angle=90,width=0.4\textwidth]{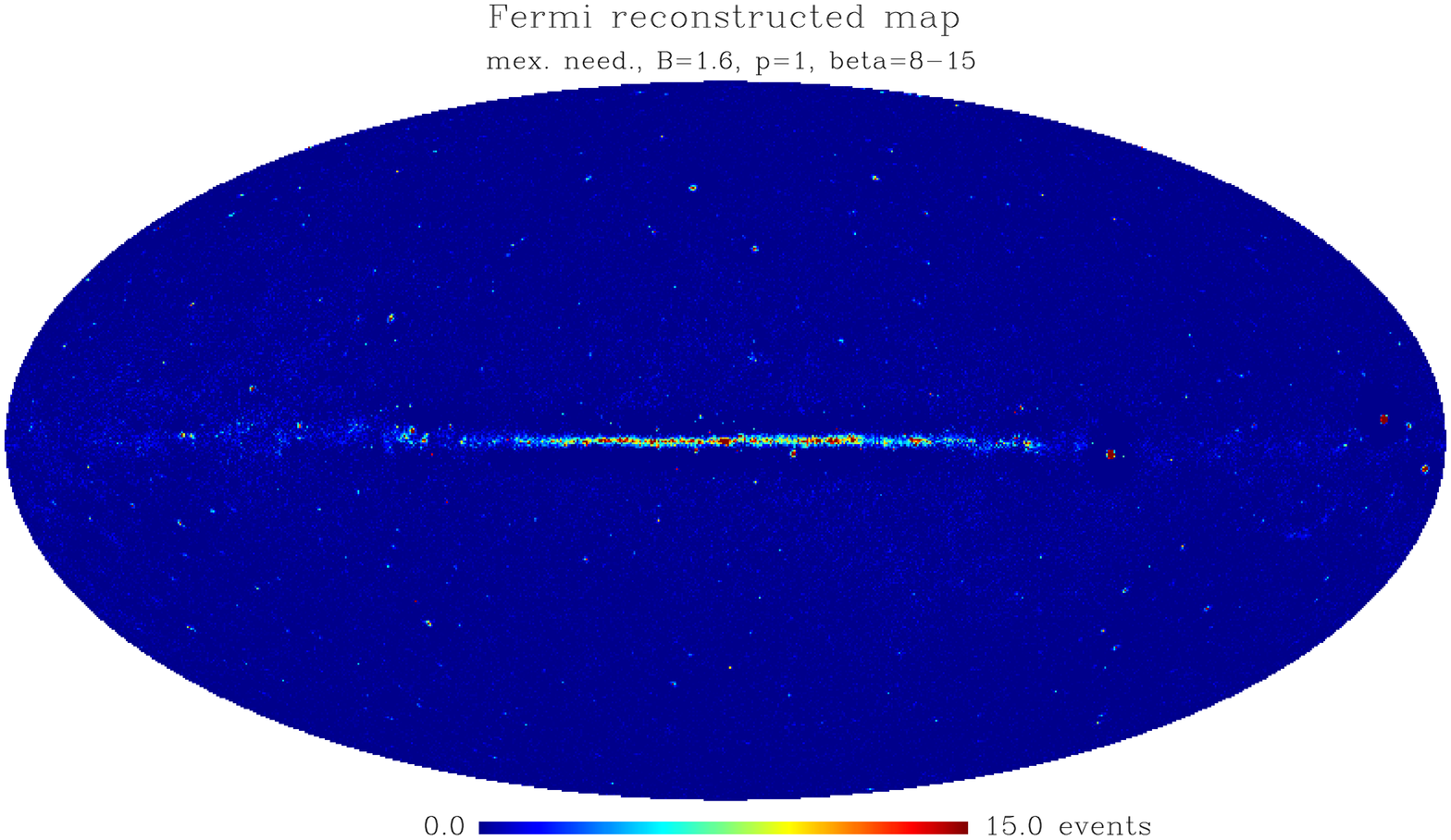} \label{fig:fermi_recon_8-15}}
  \caption{Mexican needlet backtransform of the  Fermi-LAT simulated dataset. Parameters of the transform: $B=1.6$, $p=1$. \subref{fig:fermi_recon_1-7}: used $j=1-7$. \subref{fig:fermi_recon_8-15}: used $j=8-15$. The color scales indicate the number of events reconstructed in each map (scale expanded in \subref{fig:fermi_recon_1-7}). Mollweide projections. Galactic coordinates.}
\label{fig:fermi_recon}
 \end{figure}

The figure \ref{fig:fermi_recon} reports two examples of needlet analysis applied to the Fermi-LAT simulation aforementioned. Mexican needlet transform has been applied, with $B=1.6$, $j_0=0$ and $n_j=16$; $p=1$ as usual. If we group the beta-maps in low and high multipole contributions, $j=1-7$ and $j=8-15$ respectively, figures \ref{fig:fermi_recon_1-7} and \ref{fig:fermi_recon_8-15} are obtained. The former sky map shows the diffuse emission along the galactic plane, as well as extended contributions from the most powerful point sources (Cygnus region, Vela, Crab). Therefore, needlets reveal themselves as a promising tool for investigating diffuse emissions too, treating point-like sources as noise.
On the other hand, in figure \ref{fig:fermi_recon_8-15} these sources are well visible, together with some noise at high angular frequencies, to be got rid of via thresholding algorithms (see ahead). What is most noticeable is that no trace of diffuce emission around the galactic plane is left: aside the point sources within $b=\pm5^\circ$ no events lay in the reconstructed map.

\begin{figure}[!htbp]
  \centering
  \subfigure[]{\includegraphics[width=0.3\textwidth]{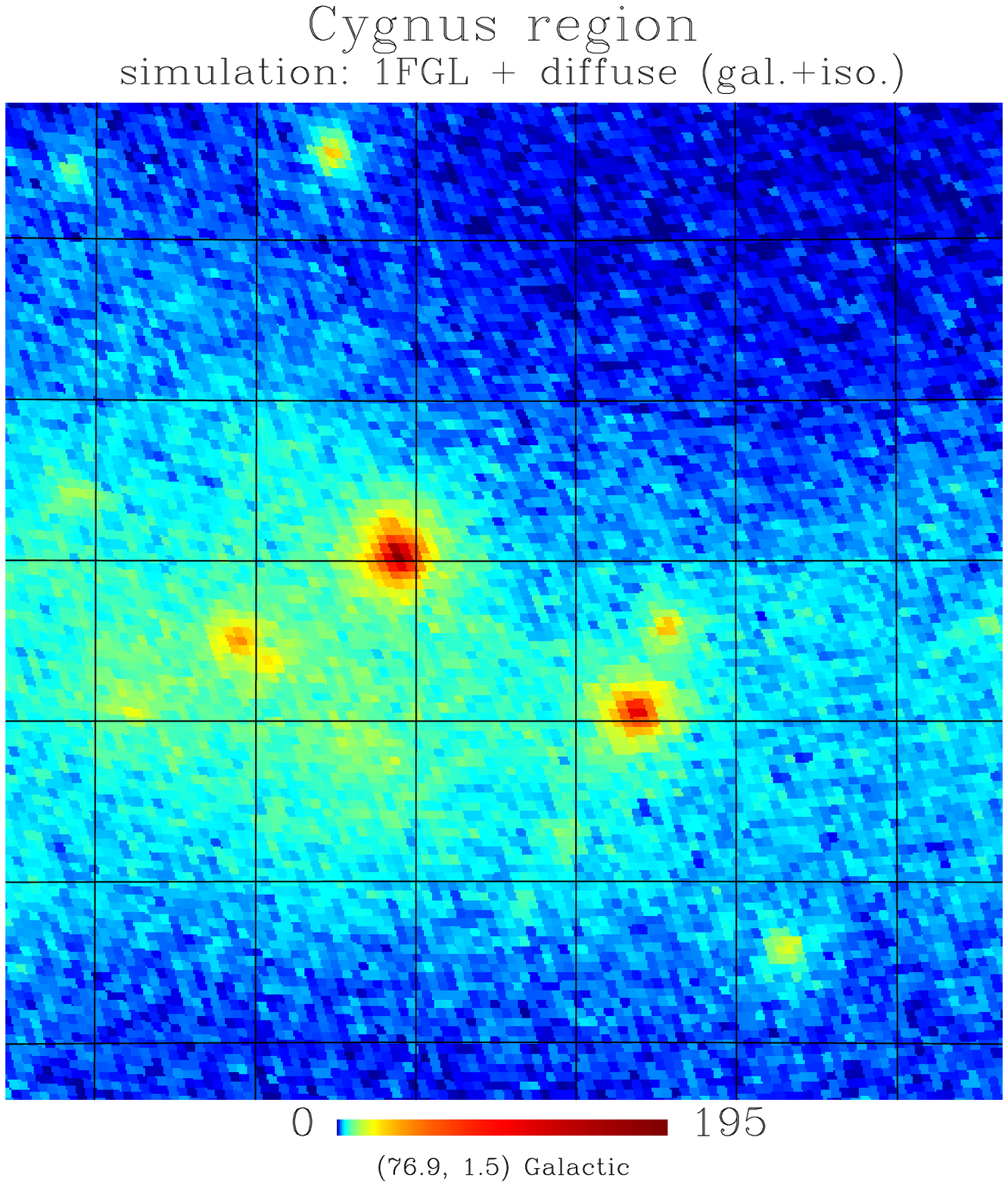} \label{fig:cygnus_map}}
  \subfigure[]{\includegraphics[width=0.3\textwidth]{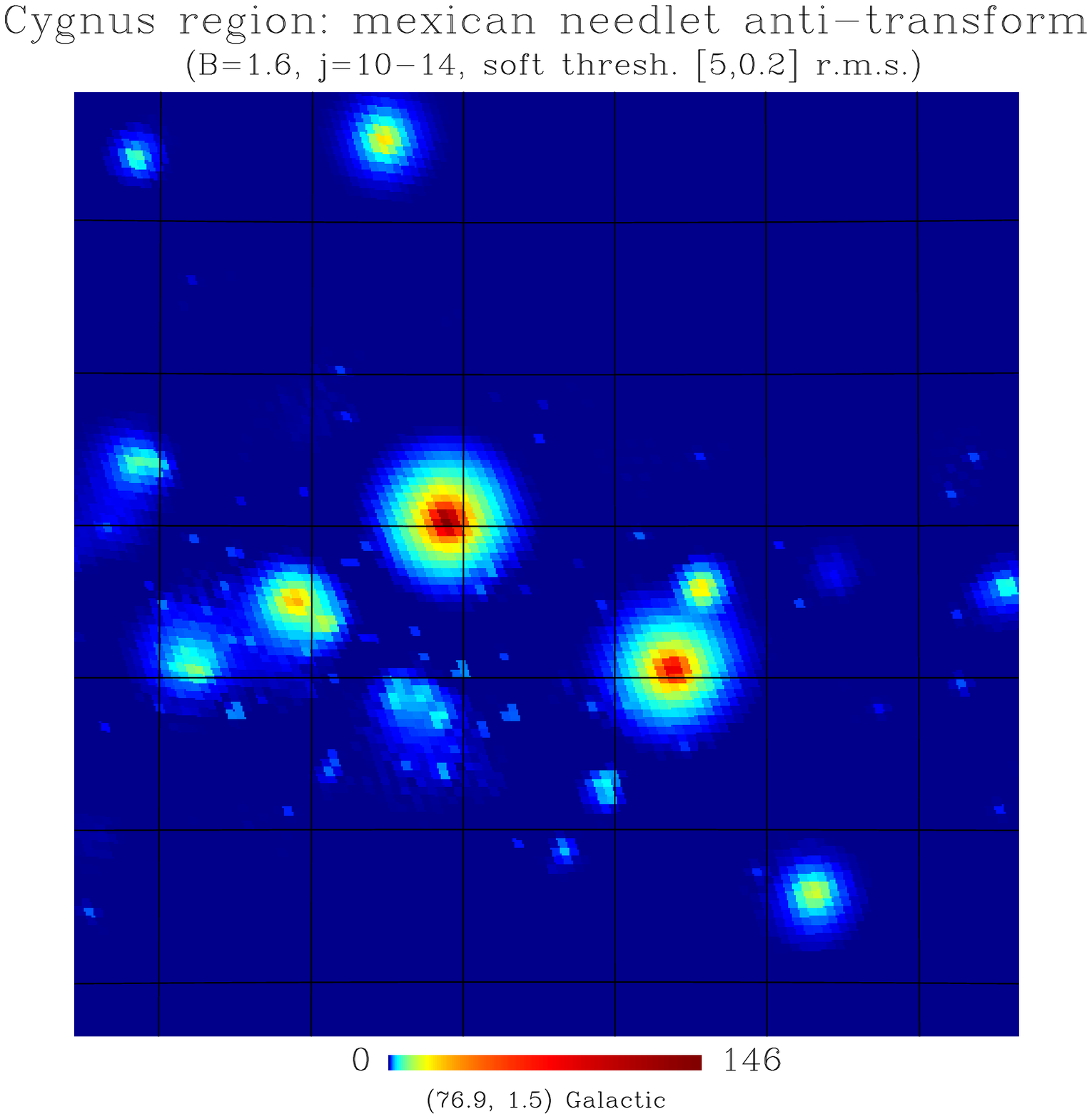} \label{fig:cygnus_antitransf}}
  \subfigure[]{\includegraphics[width=0.3\textwidth]{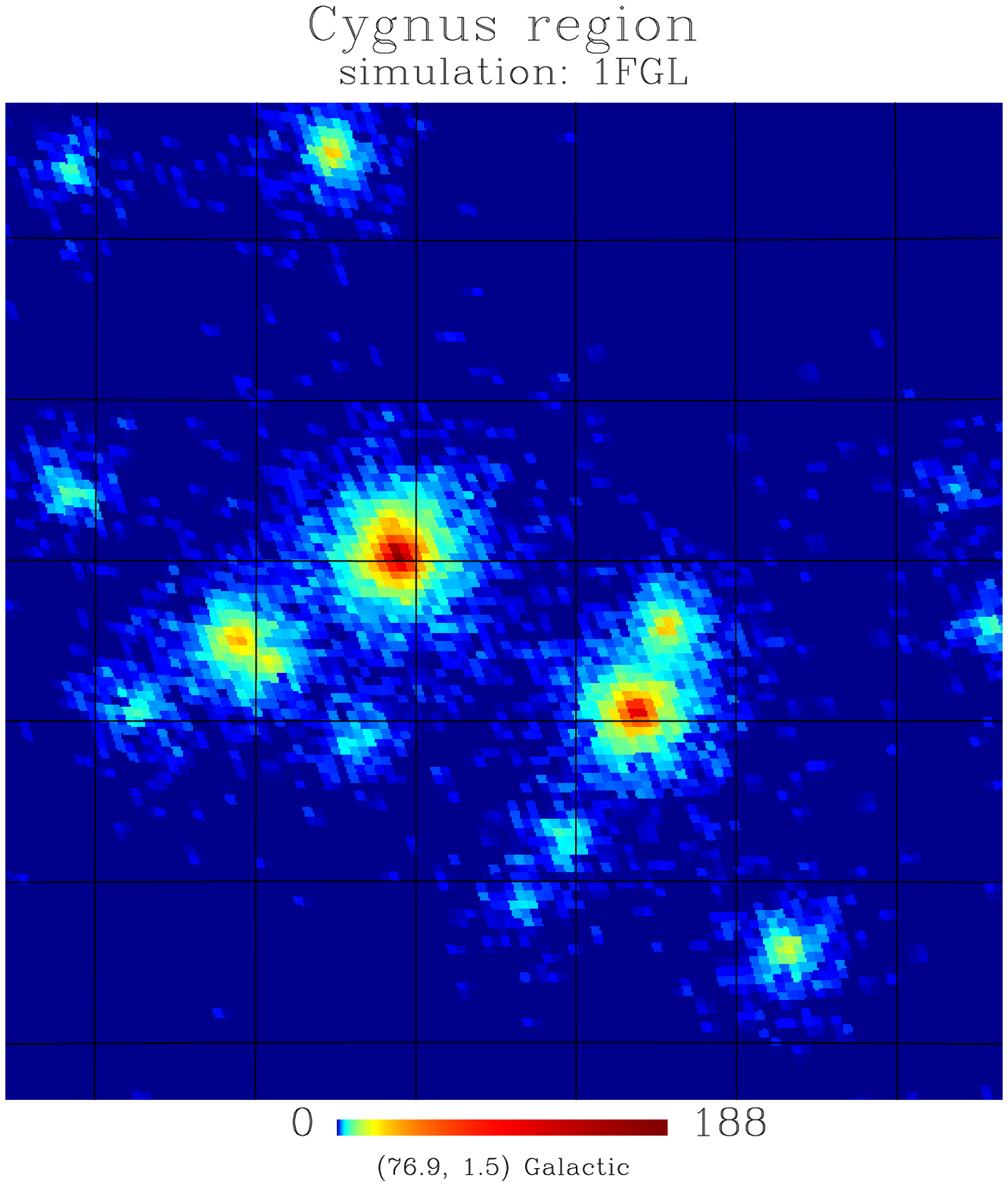} \label{fig:cygnus_sources}}
  \caption{Mexican needlet application to the Cygnus region in the simulated Fermi-LAT sky map. \subref{fig:cygnus_map}: all events simulated (point sources plus diffuse emission). \subref{fig:cygnus_antitransf}: backtransform map (used $j=9-14$, block thresholding in (ra.,de.)$=(299.5^\circ,39.0^\circ)$, radius $2^\circ$). \subref{fig:cygnus_sources}: pre-set sources. The color scales indicate the number of events. Mollweide projections. Galactic coordinates.}
 \end{figure}

To finish with, we report here one more application of the soft thresholding algorithm. It has been applied to retrieve sources in the Cygnus region, which is known to have quite a complex morphology. The figure \ref{fig:cygnus_map} reports all events that have been simulated therein: diffuse contributions and point sources. The beta-maps have been filtered estimating the average and the r.m.s. in a corner of the region under consideration (see the caption for details). The figure \ref{fig:cygnus_antitransf} reports the backtransform map realized with orders $j=9-14$ after threholding with $T=5$ and $L=0.2$. It can be appreciate how the sources appear to be well separated from each other and very few residual noise events remain.
On the other hand, the figure \ref{fig:cygnus_sources} reports the sources as they were pre-set in the simulation (i.e. without the diffuse background). Although caution is opportune because of the strong oversimplification of the analysis, the comparison of this figure to the previous one is a strong indication of how powerful the needlet approach might be in resolving point sources in crowded regions.
\section{Conclusions and perspectives}
Recently, needlets drew the attention of the scientific community for their important applications in data-analysis of cosmological data as a new form of spherical wavelets. We presented here some applications of the needlet transform to high energy cosmic ray physics, showing the very good properties of localization.The needlet transform is sensitive in the whole harmonic domain, provided that enough orders of needlets are computed. In particular, the application to a simulation of the ARGO-YBJ and the Fermi-LAT data-sets found again the well-known intermediate scale anisotropy at low orders and the point gamma-ray sources at higher orders, thus showing the possiblity of decting sources directly in the needlet space. The significance of the beta-maps is carried out quite easily if the background distribution is known and the variance of the beta coefficients may be used to threshold the signal in the beta space, which turned out to be a very promising technique.
\begin{acknowledgments}
The author wishes to thank Prof. F. K. Hansen, Prof. D. Marinucci, Prof. R. Santonico and Dr. G. Di Sciascio for fruitful discussions about needlets and potential applications in cosmic ray physics. Prof. F.K. Hansen is thanked also for his support in developing the numerical code used in this work. The author is endebted with Dr. Vincenzo Vitale for his essential help in handling Fermi-LAT simulations.
\end{acknowledgments}

\bigskip % extra skip inserted

\begin{thebibliography}{9}   % Use for  1-9  references
%\begin{thebibliography}{99} % Use for 10-99 references

%\bibitem{accelconf-ref}
%http://www.cern.ch/accelconf
%
% 1st candidate
\bibitem{icrc0507} G. Di Sciascio and R. Iuppa, Proceedings of the 32nd ICRC, 2011: 507

\bibitem{npw1} F.J. Narcowich et al., SIMA, 2006, {\bf 38}

\bibitem{gm1} D. Geller et al., Math. Z., 2009, {\bf 262}

\bibitem{mpbb08} D. Marinucci et al., MNRAS, 2008, {\bf 383}(2)

\bibitem{scodeller} S. Scodeller et al., 2010, preprint, arXiv: 1004.5576

\bibitem{bkmpAoSb} P. Baldi et al., Ann. of Stat., 2009, {\bf 37}(6A)

\bibitem{bookmarinucci} D. Marinucci and G. Peccati, Random Fields on the Sphere: Representation, Limit Theorems and Cosmological Applications, London Mathematical Society Lecture Note Series, 2011

\bibitem{needatool} D. Pietrobon et al., ApJ, 2010, {\bf 723}: 1

\bibitem{Healpix} K.M. G\`orski et al., ApJ, {\bf 622}(2)

\bibitem{icrc0518} G. Di Sciascio and R. Iuppa, Proceedings of the 32nd ICRC, 2011: 518

\bibitem{ArgoMoon} B. Bartoli et al., Physical Review D, 2011, {\bf 84}

\bibitem{ArgoBase} G. Aielli et al., NIMA, 2006, {\bf 562}

\bibitem{starckps} J.L. Starck et al., A\&A, 2009 {\bf 504}(2)

\bibitem{mrk421paper} B. Bartoli et al., ApJ, 2011 {\bf 734}

  % \bibitem{templates-ref}
%http://www.cern.ch/accelconf/templates.html

\end{thebibliography}

\end{document}